\documentclass{webofc}
\usepackage[varg]{txfonts}   % Web of Conferences font
\usepackage{hyperref}
\usepackage{url}
\hypersetup{colorlinks=true,citecolor=blue,urlcolor=blue,linkcolor=blue}
\begin{document}
\title{Interplay of prompt and non-prompt photons in photon-triggered jet observables}

\author{\firstname{Chathuranga} \lastname{Sirimanna}\inst{1}\fnsep\thanks{\email{chathuranga.sirimanna@duke.edu}}~for the JETSCAPE Collaboration 
}

\institute{Department of Physics, Duke University, Durham, NC 27708.}

\abstract{Prompt photons are important yet challenging to observe in relativistic heavy-ion collisions, as they are produced in the early stages and traverse almost the entire QGP medium without interaction. Experimental analyses typically employ isolation cuts, in the hope to identify prompt photons. Most theoretical studies consider only events with actual prompt photons, assuming no contribution from isolated non-prompt photons to reduce computational cost. For the first time, we present a study that compares simulation results generated using inclusive (bremsstrahlung) and prompt-photon events with multiple experimental observables for both $p-p$ and $Pb-Pb$ collisions at $5.02$ TeV.
Simulations are carried out using the multi-stage JETSCAPE framework tuned to describe the quenching of jets and hadrons. Isolated non-prompt photons are generated in hard photon bremsstrahlung, where the photon is radiated at a sufficient angle to the jet. Several photon triggered jet and jet substructure observables show significant contributions from inclusive photons, yielding an improvement in comparison with experimental data. Novel photon triggered jet substructure observables are also expected to show new structures, yet to be detected in experiment. This effort examines the significance of isolated non-prompt photons using parameters tuned for a simultaneous description of the leading hadron and jet spectrum, and thus provides an independent verification of the multistage evolution framework.
}
\maketitle

\vspace{-0.5cm}

\section{Introduction}
\label{sec:intro}

Relativistic heavy-ion collisions are used to study the hot, dense state of deconfined quarks and gluons, known as the Quark-Gluon Plasma (QGP), believed to have existed shortly after the Big Bang. Hard probes, such as jets and prompt photons, are valuable tools for investigating the properties of the QGP. Since hard probes interact with the QGP as they traverse it, they carry imprints of its characteristics, such as temperature, density, and transport properties. Among these, prompt photons are particularly important because they are produced directly in hard subprocesses and do not interact with the QGP medium \cite{Wang:1996yh}. However, experimental studies involving prompt photons are challenging, as their production is rare and distinguishing them from photons originating in parton showers or hadronic decays is difficult.

The most common method for identifying prompt photons is the isolation cut, where the total energy of particles surrounding the photon within a cone of defined radius must be less than a specified threshold \cite{ATLAS:2023iad, CMS:2017ehl, ATLAS:2018dgb}. While the majority of photons identified using this method are indeed prompt, there can still be a significant contribution from non-prompt photons—those not produced in leading-order hard scattering processes. In these proceedings, we perform a systematic study of photon-triggered jets and the contribution of non-prompt photons to the isolated photons. Since many theoretical simulations consider only events with photons originating from hard scatterings when studying photon-triggered jets, understanding the role of non-prompt photons is crucial.

\section{Simulating photon-triggered jets}
\label{sec:model}

In this study, the \textsc{JETSCAPE} framework \cite{Putschke:2019yrg, JETSCAPE:2024nkj, JETSCAPE:2025rip} is used to generate both high-statistics inclusive events and prompt photon events, the latter consisting exclusively of events where photons are produced in the initial hard scattering. To simulate the bulk medium in $\sqrt{s_{NN}} = 5.02~\text{TeV}$ Pb–Pb collisions, we use pre-generated event-by-event 2+1D \textsc{VISHNU} \cite{Shen:2014vra} hydrodynamic profiles with \textsc{TRENTo} \cite{Moreland:2014oya} 2D initial conditions. The initial hard scattering is simulated using \textsc{PYTHIA}, and all partons are passed through energy loss modules to simulate the intermediate partonic shower. In $p$–$p$ simulations, the \textsc{MATTER} \cite{Majumder:2013re, Cao:2017qpx} vacuum shower is used with the PP19 tune \cite{JETSCAPE:2019udz}, while in Pb–Pb simulations, a \textsc{MATTER}+\textsc{LBT} \cite{Wang:2013cia, He:2015pra, Cao:2016gvr} module combination is used with the AA22 tune \cite{JETSCAPE:2022jer} for multistage evolution. After the partonic shower, all final-state partons are hadronized using the \textsc{PYTHIA} Lund String Model.

\section{Results and discussion}
\label{sec:results}

We use the \textsc{JETSCAPE} framework to explore photon-triggered jets. In this study, we investigate the contribution of non-prompt photons to the isolated photon sample by employing two types of event selections: total inclusive events (also referred to as full events) and prompt-photon events, which include only events where photons are produced in the initial hard scattering. In these proceedings, we compare the results for photon-triggered jet transverse momentum imbalance (also known as $\gamma$-jet asymmetry) and azimuthal correlations, generated using both full and prompt-photon event simulations, with available experimental data.

\vspace{-0.2cm}

\subsection{Photon-jet transverse momentum imbalance }

In a rare hard scattering event, where a hard photon is created with large transverse momentum, a jet (usually a quark jet) with matching transverse momentum is produced in the direction opposite to the photon, due to momentum conservation. $\gamma$-jet asymmetry refers to the transverse momentum imbalance between the prompt photon and the jet that recoils against it in a high-energy collision. Since the photon does not interact strongly with the medium, it provides a clean reference for the initial momentum of the associated jet. The asymmetry is typically quantified by the ratio $x_{J\gamma} = \frac{p_T^{\text{jet}}}{p_T^\gamma}$, where $p_T^{\text{jet}}$ and $p_T^\gamma$ are the transverse momenta of the jet and photon, respectively. In the absence of medium effects, $x_{J\gamma}$ would peak near 1, indicating balanced momentum. However, in a medium like the QGP, energy loss by the jet causes a shift toward lower values of $x_{J\gamma}$, revealing information about jet quenching.

\begin{figure*}[htb!]
\centering
\sidecaption
\includegraphics[width=0.66\textwidth]{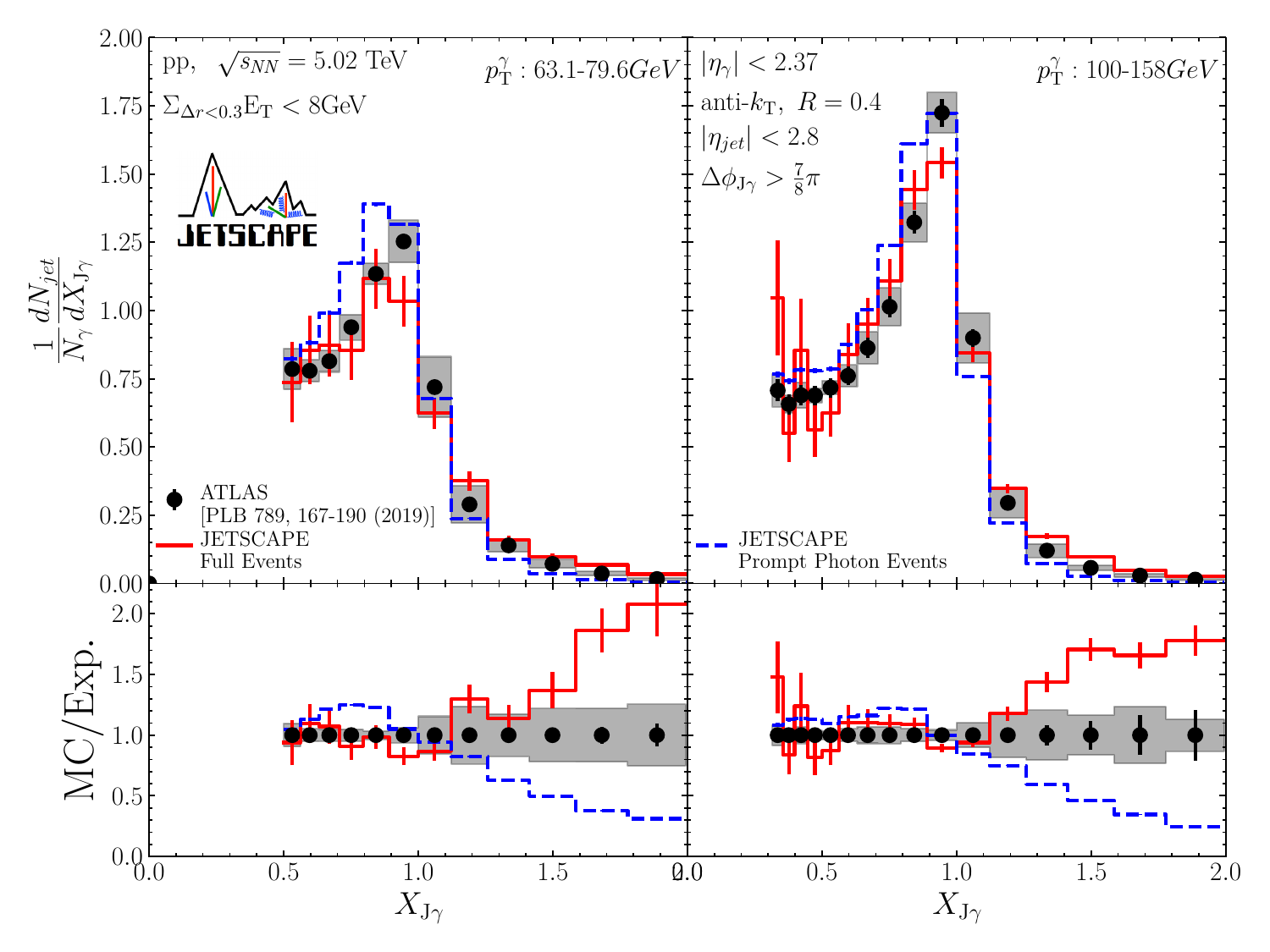}
\caption{Photon-triggered jet asymmetry in $p-p$ collisions at $\sqrt{s_{NN}} = 5.02~\text{TeV}$, compared with ATLAS experimental results \cite{ATLAS:2018dgb}. The solid red lines represent the total inclusive events, while the dashed blue lines represent the prompt photon events. The bottom panel shows the ratio of \textsc{JETSCAPE} results to the experimental results. }
\label{fig:Xj_ATLAS_pp}
\end{figure*}

\begin{figure*}[htb!]
\centering
\sidecaption
\includegraphics[width=0.66\textwidth]{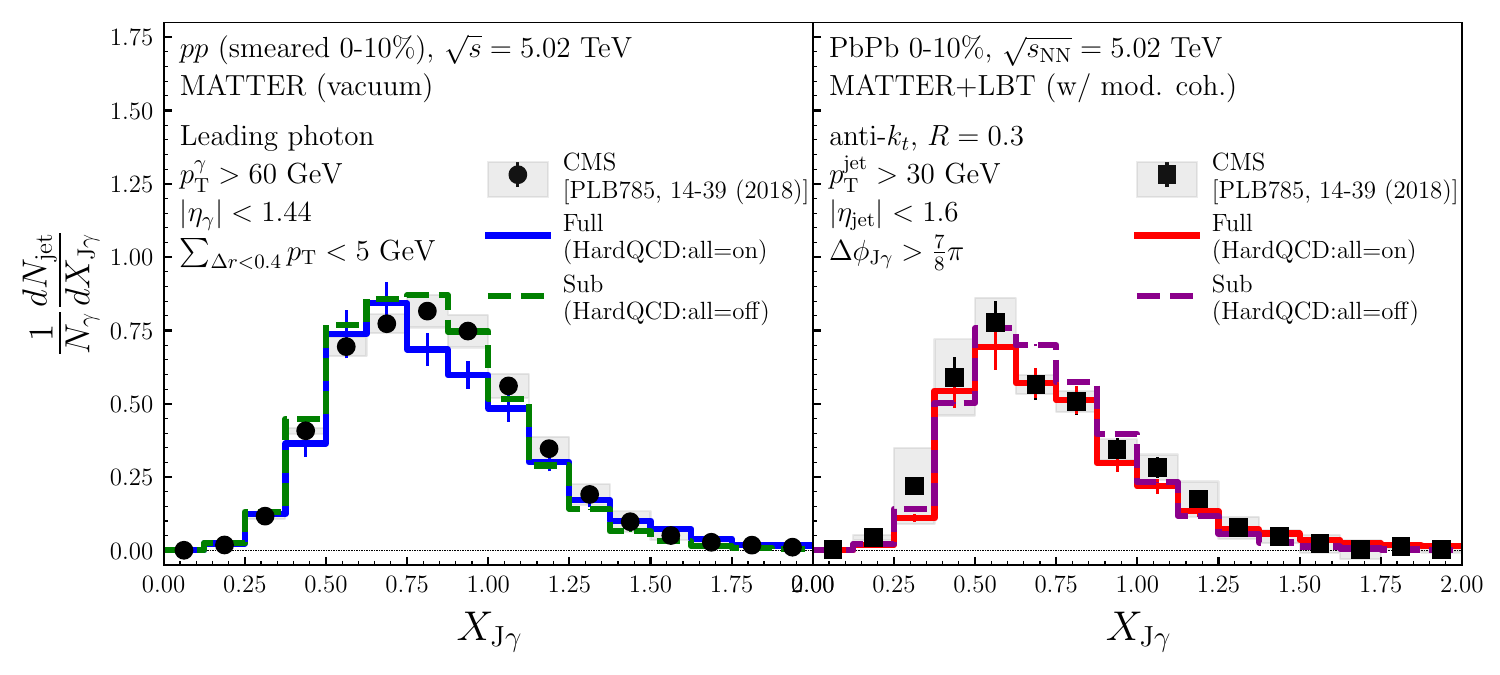}
\caption{Photon-triggered jet asymmetry in $p-p$ and $0-10\% ~ Pb–Pb$ collisions at $\sqrt{s_{NN}} = 5.02~\text{TeV}$, compared with CMS experimental results \cite{CMS:2017ehl}. The solid and dashed lines represent total inclusive and prompt photon events, respectively.}
\label{fig:Xj_CMS}
\end{figure*}

As shown in Figures \ref{fig:Xj_ATLAS_pp} and \ref{fig:Xj_CMS}, \textsc{JETSCAPE} results for both full events and prompt photon events show excellent agreement with experimental results at small $x_{J\gamma}$ values. However, at large $x_{J\gamma}$ values, a noticeable difference is observed for both $p-p$ and $Pb-Pb$ collisions. The statistical errors associated with full events are significantly larger compared to those for prompt photon events. This is primarily because observing an event with a prompt (or isolated) photon is rare, as the production cross-section for prompt photons is suppressed by the fine-structure constant, $\alpha \sim \frac{1}{137}$. Additionally, isolated non-prompt photons are primarily responsible for the large difference at large $x_{J\gamma}$ values, which is clearly visible in the ratio plots shown in the bottom panel of Figure \ref{fig:Xj_ATLAS_pp}. Since isolated non-prompt photons are produced in the intermediate partonic shower or hadronic splittings, the away-side jet typically carries much larger transverse momentum compared to the photon.

Furthermore, as clearly shown in Figure \ref{fig:Xj_CMS}, the distribution peaks around $x_{J\gamma} = 1$ for $p-p$ collisions, while the peak shifts toward smaller $x_{J\gamma}$ values for $Pb-Pb$ collisions. The $p-p$ distribution is also slightly skewed toward smaller $x_{J\gamma}$ values, as these results are smeared using the smearing function for $0-10\%$ centrality.

\vspace{-0.2cm}

\subsection{Photon-jet azimuthal correlation }

The $\gamma$-jet azimuthal correlation measures the angular distribution between a prompt photon and the associated jet in the plane transverse to the beam direction. In a hard scattering event, the photon and jet are initially produced nearly back-to-back, with an azimuthal angle difference $\Delta\phi \sim \pi$, due to momentum conservation. Since the photon does not interact strongly with the medium, it preserves information about the initial scattering, while the jet may experience energy loss and angular broadening if it traverses the QGP medium. Studying the $\gamma$-jet azimuthal correlation thus provides insight into medium-induced effects, such as transverse momentum broadening.

\begin{figure*}[htb!]
\centering
\includegraphics[width=\textwidth]{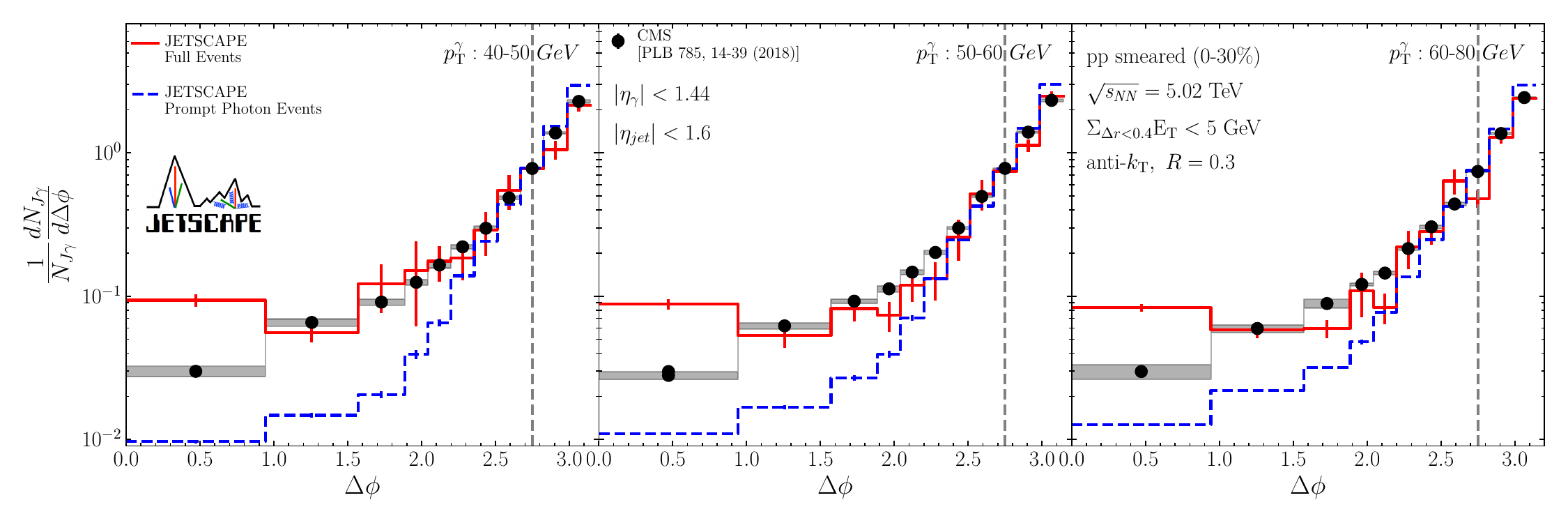}
\caption{Photon-triggered jet azimuthal correlation in $p-p$ collisions at $\sqrt{s_{NN}} = 5.02~\text{TeV}$, compared with CMS experimental results \cite{CMS:2017ehl}. The solid red lines represent the total inclusive events, while the dashed blue lines represent the prompt photon events.}
\label{fig:dPhi_pp}
\end{figure*}

In Figure \ref{fig:dPhi_pp}, \textsc{JETSCAPE} results for $p-p$ collisions at $\sqrt{s_{NN}} = 5.02~\text{TeV}$ are compared with CMS results. It is clear that the full events, despite large statistical errors, describe the entire range of the experimental data well, except for the smallest $\Delta\phi$ bin across all $p_T^\gamma$ regions. In contrast, results generated using prompt photon events fail to reproduce the experimental data at smaller $\Delta\phi$ values. The difference between prompt photon and full events becomes small when $\Delta\phi > \frac{7\pi}{8}$. Since an azimuthal cut of $\Delta\phi > \frac{7\pi}{8}$ was applied when generating the $x_{J\gamma}$ distribution to ensure the isolated photon and jet are approximately back-to-back, this could explain why there is no significant difference between full and prompt photon events in Figures \ref{fig:Xj_ATLAS_pp} and \ref{fig:Xj_CMS}.

\begin{figure*}[htb!]
\centering
\includegraphics[width=\textwidth]{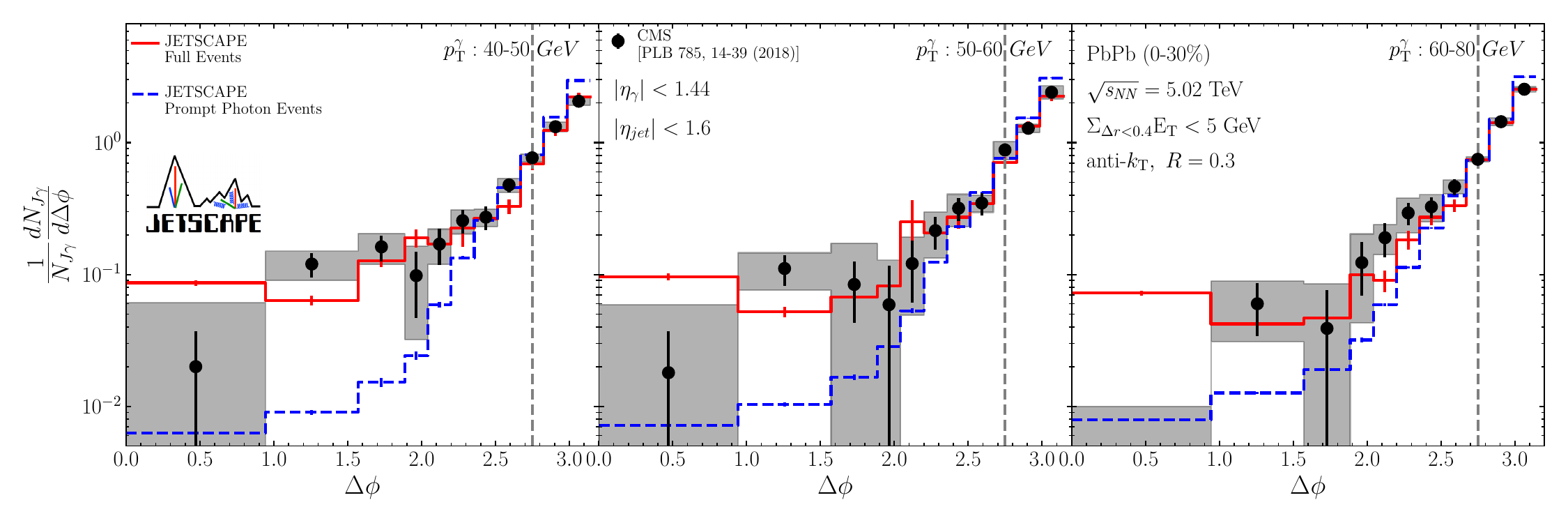}
\caption{Photon-triggered jet azimuthal correlation in $0-30\% ~ Pb-Pb$ collisions at $\sqrt{s_{NN}} = 5.02~\text{TeV}$, compared with CMS experimental results \cite{CMS:2017ehl}. The solid red lines represent the total inclusive events, while the dashed blue lines represent the prompt photon events.}
\label{fig:dPhi_PbPb}
\end{figure*}

Figure \ref{fig:dPhi_PbPb} shows the comparison of \textsc{JETSCAPE} results for $Pb-Pb$ collisions at $\sqrt{s_{NN}} = 5.02~\text{TeV}$ with CMS results. A similar trend to that observed in Figure \ref{fig:dPhi_pp} for $p-p$ collisions is seen here. Since the experimental results have significantly large statistical and systematic uncertainties, both the full and prompt photon events describe the data within the error margins. However, the full events provide a particularly good description, similar to what was observed for the $p-p$ results.

\section*{ACKNOWLEDGMENTS}

This work was supported in part by the National Science Foundation (NSF) within the framework of the JETSCAPE collaboration, under grant number OAC-2004571 (CSSI:X-SCAPE), and in part by the US Department of Energy, Office of Science, Office of Nuclear Physics under grant numbers \rm{DE-SC0013460} and \rm{DE-FG02-05ER41367}.

%
% BibTeX or Biber users please use (the style is already called in the class, ensure that the "woc.bst" style is in your local directory)
\bibliography{photonjet} % Replace "your_bib_file" with the actual name of your .bib file

\end{document}